\documentclass{article}

     \usepackage[final,nonatbib]{neurips_2019}

\usepackage[utf8]{inputenc} 
\usepackage[T1]{fontenc}    
\usepackage{hyperref}       
\usepackage{url}            
\usepackage{booktabs}       
\usepackage{amsfonts}       
\usepackage{nicefrac}       
\usepackage{microtype}      

\usepackage{graphicx}
\usepackage{amsmath}
\usepackage{fullpage}
\usepackage{multirow}
\usepackage{enumitem}
\usepackage{comment}
\usepackage{wrapfig}

\title{A fully 3D multi-path convolutional neural network with feature fusion and feature weighting for automatic lesion identification in brain MRI images}

\author{%
 Yunzhe Xue\\
  Department of Computer Science\\
  New Jersey Institute of Technology\\
  Newark, NJ, USA\\
  \texttt{yx277@njit.edu} \\
   \And
  Meiyan Xie \\
  Department of Computer Science\\
  New Jersey Institute of Technology\\
  Newark, NJ, USA\\
  \texttt{mx42@njit.edu} \\
   \AND
  Fadi G. Farhat \\
  Department of Computer Science\\
  New Jersey Institute of Technology\\
  Newark, NJ, USA\\
  \texttt{fgf4@njit.edu} \\
\And
  Olga Boukrina \\
 Stroke Rehabilitation Research\\
  Kessler Foundation\\
  West Orange, NJ, USA\\
  \texttt{oboukrina@kesslerfoundation.org} \\
\And
  A. M. Barrett \\
  Emory University and Atlanta VA Medical Center\\
  Atlanta, GA, USA\\
  \texttt{ambarrettmd@gmail.com} \\
\And
  Jeffrey R. Binder \\
  Department of Neurology\\
  Medical College of Wisconsin\\
  Milwaukee, WI, USA\\
  \texttt{jbinder@mcw.edu} \\
\And
Usman W. Roshan \\
  Department of Computer Science\\
  New Jersey Institute of Technology\\
  Newark, NJ, USA\\
  \texttt{usman@njit.edu} \\
\And
William W. Graves \\
 Department of Psychology\\
 Rutgers University - Newark\\
Newark, NJ, USA\\
 \texttt{william.graves@rutgers.edu} 
}

\begin{document}

\maketitle

\begin{abstract}
We propose a fully 3D multi-path convolutional network to predict stroke lesions from 3D brain MRI images. Our multi-path model has independent encoders for different modalities containing residual convolutional blocks, weighted multi-path feature fusion from different modalities, and weighted fusion modules to combine encoder and decoder features. 
Compared to existing 3D CNNs like DeepMedic, 3D U-Net, and AnatomyNet, our networks achieves the highest statistically significant cross-validation accuracy of 60.5\% on the large ATLAS benchmark of 220 patients. We also test our model on multi-modal images from the Kessler Foundation and Medical College Wisconsin and achieve a statistically significant cross-validation accuracy of 65\%, significantly outperforming the multi-modal 3D U-Net and DeepMedic. Overall our model offers a principled, extensible multi-path approach that outperforms multi-channel alternatives and achieves high Dice accuracies on existing benchmarks.
\end{abstract}

\section*{Methods}
\begin{figure}[h]
  \centering
  \includegraphics[scale=.5]{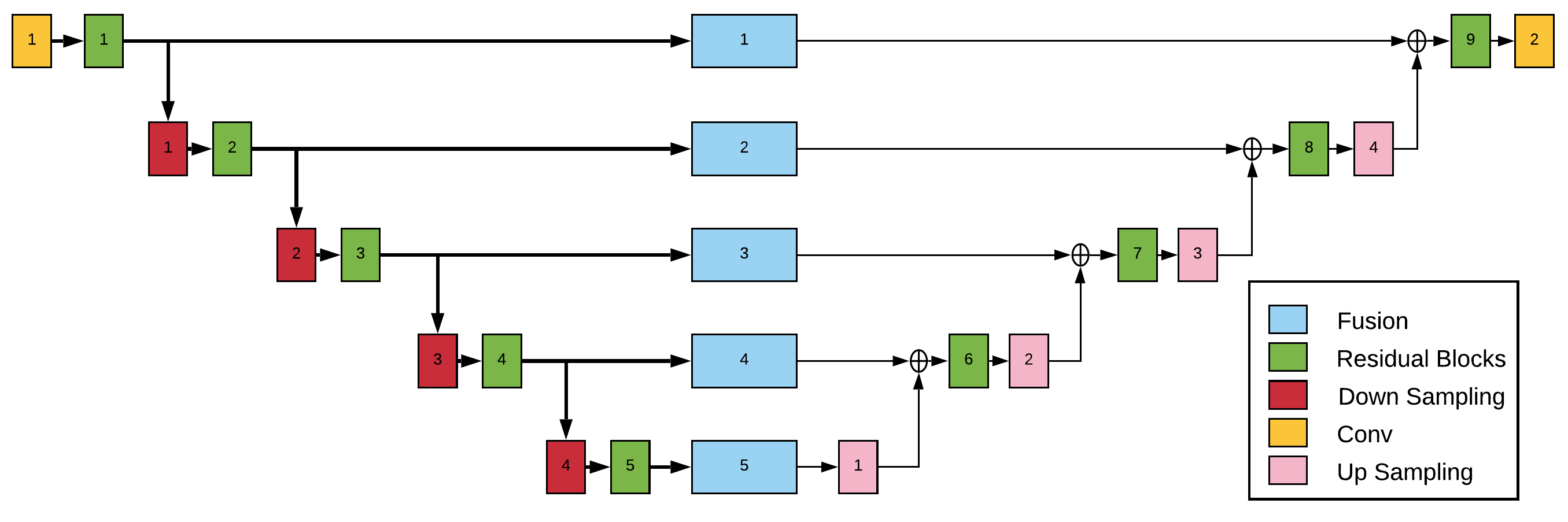} \\
  \caption{Overview of our 3D convolutional neural net system. \label{overview}}
\end{figure}

\vspace{-.1in}
\paragraph{Fully 3D multi-path convolutional neural network}
Our contribution is a fully 3D convolutional network for predicting stroke lesions in brain MRI images. Our model contains only 3D convolutional kernels, 3D feature fusion modules, and feature weighting methods. We use separate encoders for each modality which we then fuse in a custom designed feature fusion module. The fused features are then added to outputs from the upsampling blocks. We use squeeze-and-excitation to weigh channels for different modalities as well as a simple custom designed amplified weighting designed to fit small lesions that are hard to detect. Our overall model structure is given in Figure~\ref{overview}. The U-shaped network with encoders, decoders, and connections between them is a widely used structure for segmentation problems \cite{ronneberger2015u}. 

\subsubsection*{Variants of our model}
{\bf Single path:} Here we have a single encoder combined for all modalities. In this version different modalities are stacked as multiple channels. With this variant we can see the effect of a multi-channel vs. multi-path approach for different input modalities.\\
{\bf No fusion:} This applies only to ATLAS data where we have single modality T1 images. Since the images have just one modality there are no features to fuse. Thus we we replace the fusion module with identity mapping as in the original 3D UNet.\\
{\bf Basic:} This is our basic multi-path model with the basic fusion component.\\
{\bf Squeeze-and-excitation in fusion (SE):} Here we consider the fusion block with the squeeze-and-excitation component \cite{hu2018squeeze} in fusion to weigh channels across different modalities. This is a kind of attention mechanism \cite{vaswani2017attention} focusing on channels. The idea is to use pooling to compress features to a point followed by fully connected layers and non-linear activation (sort of a mini encoder decoder setup). The sigmoid layer outputs a number between 0 and 1 that is used to weight the channel.\\
{\bf SE and Amplified Weighting (SE+AW):} During our experimental design we found that small lesions are particularly hard to map. In fact our ATLAS benchmark has about a quarter of images with lesion volumes below $1000mm^3$ and a third quarter below $10000mm^3$. More interestingly we found 25 samples in ATLAS which give a 0 Dice coefficient during training. We call these \emph{hard} samples and found their gradient values in the residual block in the decoder to be much smaller compared to that of other samples (about one thousand times). One explanation for this is the addition operation between features from the upsampler and the fusion module. Upon closer examination we find the output from the fusion to have a much larger value than the upsampler before entering the residual block (30 to 100 times). As a possible fix we simply amplify the weights from the upsampler by a factor of 10. We call this Amplified Weighting (AW).\\
{\bf ReLUOnly:} Instead of amplifying the upsampler weights as above we consider an alternative approach to balance the magnitude of outputs from the fusion and upsampler modules. In our residual block we move the first normalization layer to the end as the final layer just before addition. This is known as the ReLU-Only pre-activation \cite{he2016identity}. With this modification the outputs from the residual blocks get normalized to between 0 and 1. These are then given as inputs to the fusion module and because they are normalized the outputs from the fusion module are not much larger than the upsampler outputs (about 10 to 30 times). In fact the squeeze-and-excitation module that assigns weights to different channels is partly responsible for the weight inflation. This variant includes the SE variant above.


\subsection*{Data, other methods, measure of accuracy and statistical significance}
{\bf Imaging Data:} We obtained high-resolution (1 mm$^{3}$) whole-brain MRI scans from 25 patients from the Kessler Foundation (KES), a neuro-rehabilitation facility in West Orange, New Jersey. We also obtained 20 high-resolution scans from the Medical College of Wisconsin (MCW). We included subacute ($<$ 5 weeks post stroke) and chronic ($>$ 3 months post stroke) cases. Strokes of both hemorrhagic and ischemic etiology were included. 
We also obtained 220 high resolution scans from the public ATLAS database \cite{liew2018large}. In all datasets hand-segmented lesions by trained human experts are also available. \\
{\bf Comparison of CNN Methods:} We compared our CNN to three state of the art recently published CNNs. 
(1)  DeepMedic \cite{kamnitsas2017efficient}: This is a popular dual-path 3D convolutional neural network with a conditional random field to account for spatial order of slices. (2) AnatomyNet \cite{zhu2019anatomynet}: This is a convolutional neural network with residual connections \cite{he2016deep}
and squeeze-excitation blocks \cite{hu2018squeeze}. (3) 3D-UNet \cite{ronneberger2015u}: A 3D U-Net that obtained third place in the BRATS 2017 multimodal brain tumor segmentation challenge. \\
{\bf Measure of accuracy: Dice coefficient:} The Dice coefficient is typically used to measure the accuracy of predicted lesions in MRI images \cite{zijdenbos1994morphometric}. Starting with the human binary mask as ground truth, each predicted voxel is determined to be either a true positive (TP, also one in true mask), false positive (FP, predicted as one but zero in the true mask), or false negative (FN, predicted as zero but one in the true mask). The Dice coefficient is formally defined as $DICE=\frac{2TP}{2TP+FP+FN}$.\\
{\bf Measure of statistical significance: Wilcoxon rank sum test:} The Wilcoxon rank sum test \cite{wilcoxon1945individual} (also known as the Mann-Whitney U test) can be used to determine whether the difference between two sets of measurements (methods in our case) is significant. While the t-test has been shown to have high Type 1 error possibly due to violation of independence in the folds \cite{dietterich1998approximate}, the Wilcoxon rank test has been previously recommended for testing across datasets \cite{demvsar2006statistical} and also applied on a single dataset \cite{japkowicz2011evaluating} like we do. 

\section*{Results}
\vspace{-.1in}
We perform a five-fold cross-validation. On KES+MCW our splits are chosen randomly whereas in ATLAS we first rank the samples by their training accuracy in increasing order. We then group every five together and rotate the test sample by picking the first from each group of five, then the second and so forth. This gives us five sets of train and test samples. Our goal here is to include hard to train samples in the training set so as to better predict hard ones in the test. We apply all methods to the same set of splits in both datasets. Each of our MRI images is aligned to standard space. 
We normalized each 3D image by subtracting the mean and dividing by variance of the image's pixel values. 

\begin{figure}[h]
  \centering
\begin{tabular}{cc}
  \includegraphics[scale=.3]{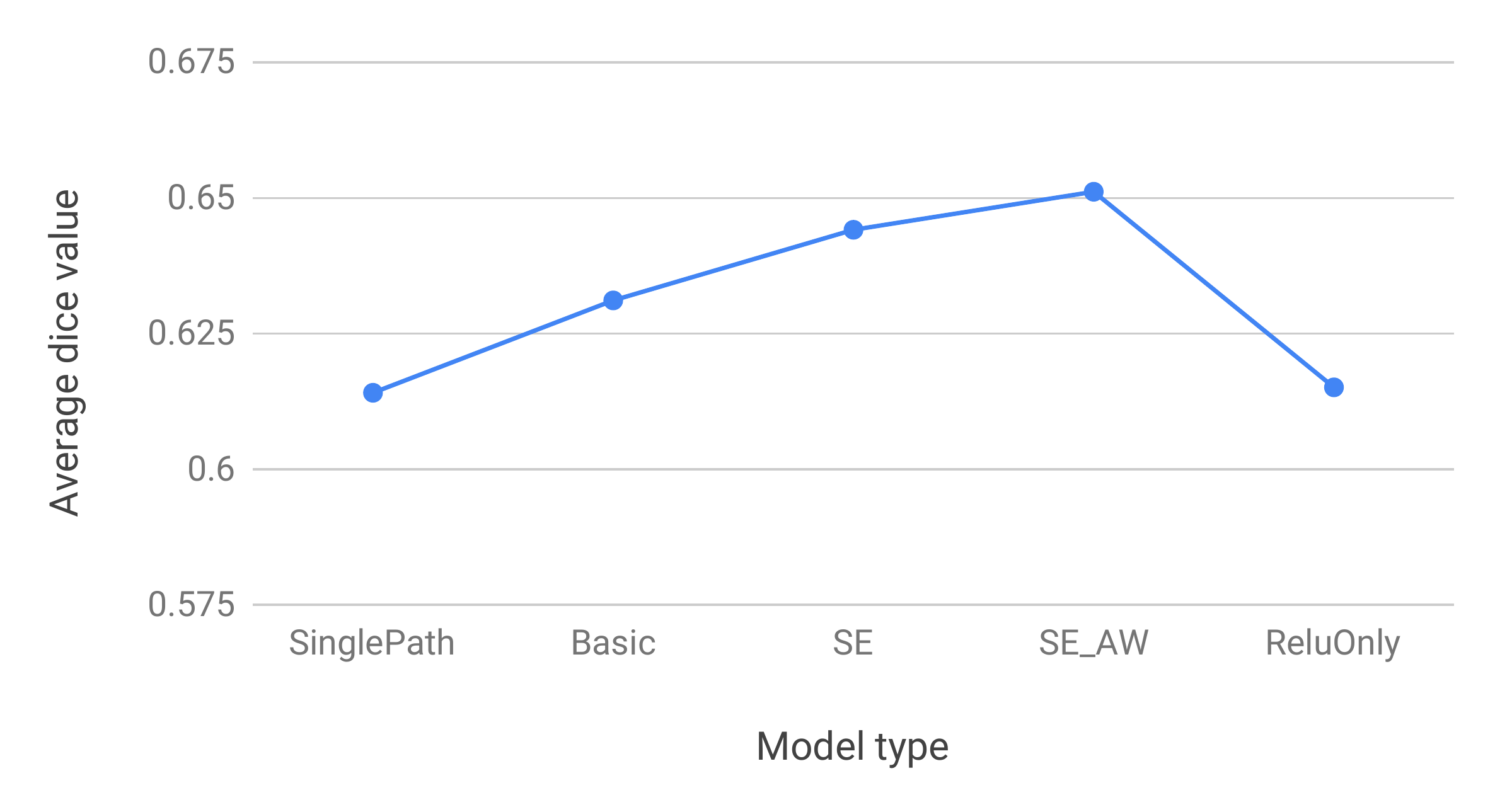} &  \includegraphics[scale=.22]{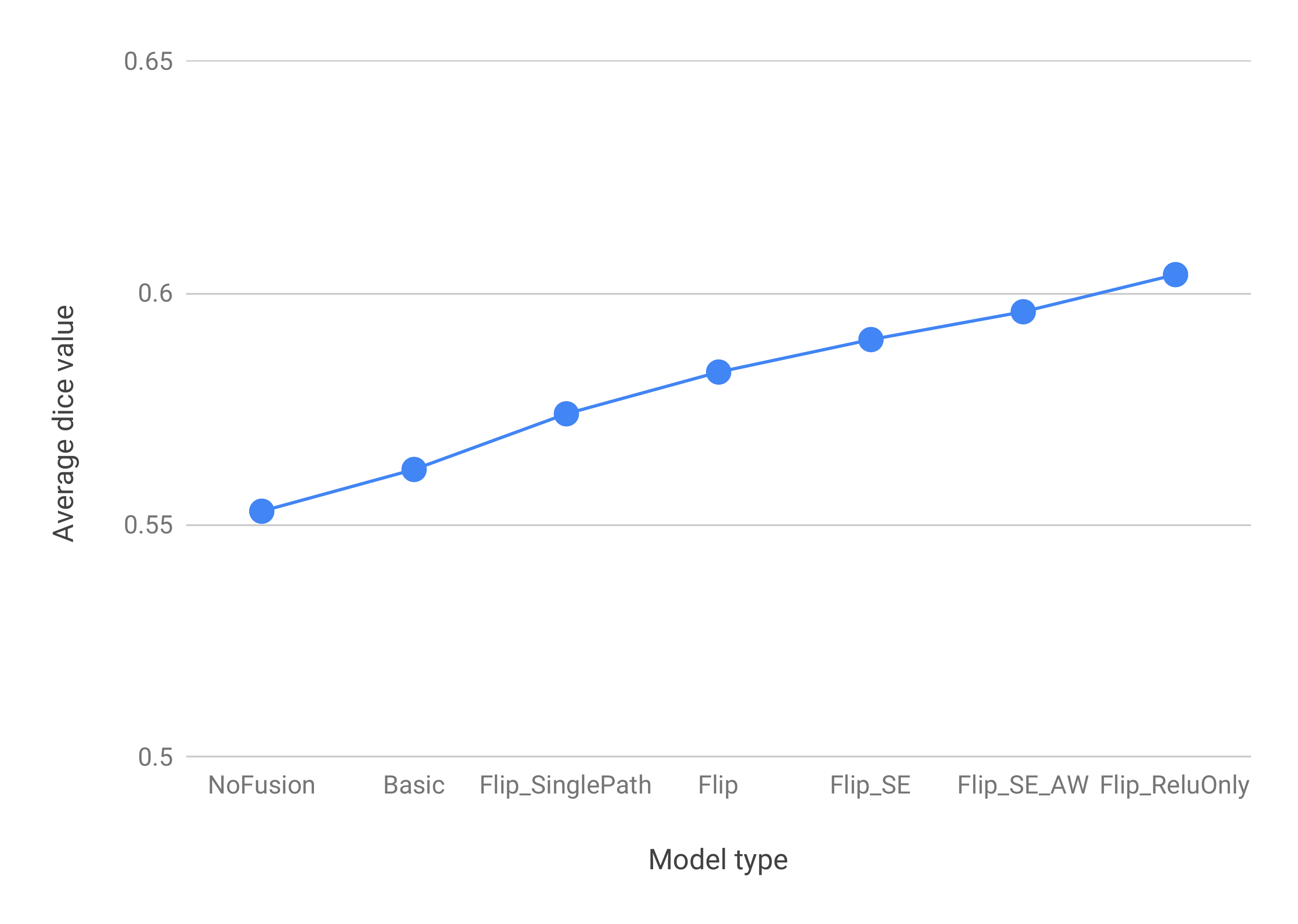}  \\
  (a) Kessler+MCW & (b) ATLAS \\
  \end{tabular}
  \caption{Average five-fold test Dice values of variants of our model on KES+MCW and ATLAS datasets. \label{results1}}
\end{figure}

In Figure~\ref{results1} we compare the average Dice coefficients of test samples of each of our variants. On KES+MCW (Figure~\ref{results1}(a)) we see that our Basic variant (multi-path approach) has about a 2\% improvement over the SinglePath variant that stacks different modalities as multiple channels. Thus treating each modality independently with their own encoders has a noticeable advantage. We see the same in ATLAS (Figure~\ref{results1}(b)) if we consider the flipped image as a second modality. 

In fact in ATLAS we see that using the flipped image as another modality brings a considerable improvement over using just the single T1 images alone. The Flip variant in Figure~\ref{results1}(b) has 58.4\% Dice value whereas the Basic variant is at 56.3\% (with the difference having a p-value of .086). 
Following the basic model we see that both squeeze and excite blocks and the amplified weighting improves overall test accuracy on both datasets. Thus upweighting features from the upsampler (as in our AW variant) help to improve not just training accuracy on hard samples but also the overall test accuracy. 

\begin{figure}[h]
  \centering
\begin{tabular}{cc}
  \includegraphics[trim=0 20 0 20, clip,scale=.37]{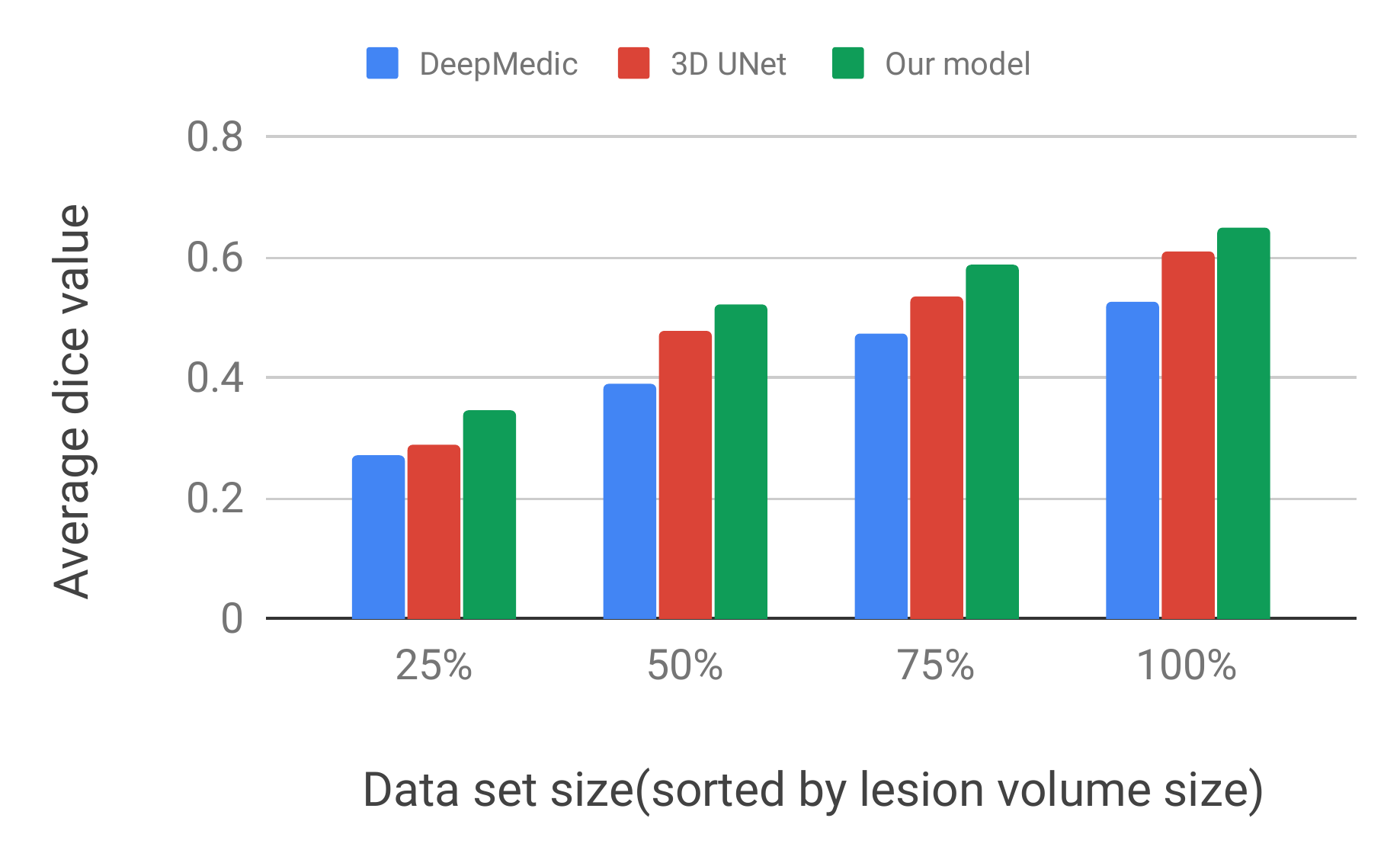} &  \includegraphics[trim=0 20 0 20, clip,scale=.37]{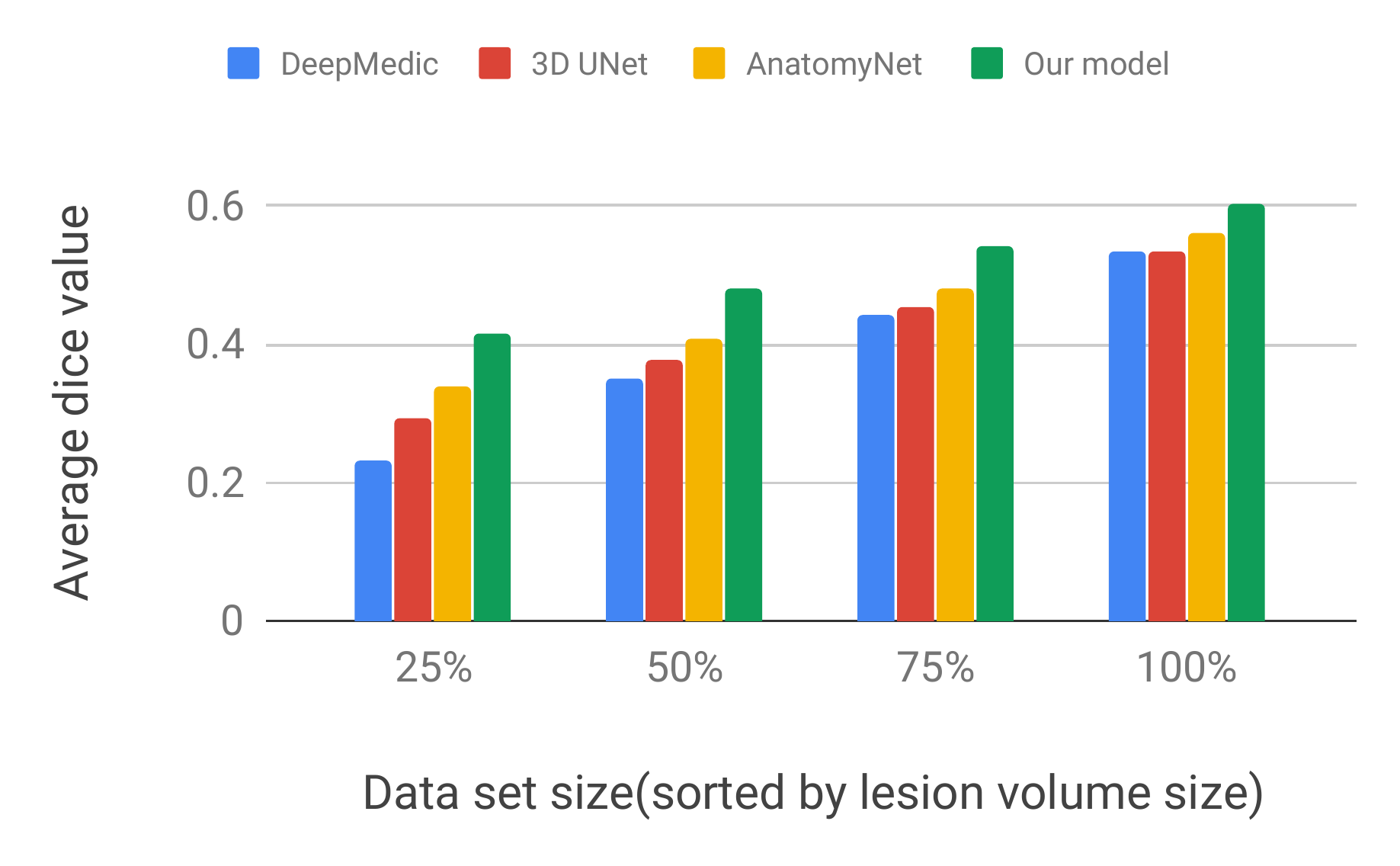}  \\
  (a) Kessler+MCW & (b) ATLAS \\
  \end{tabular}
  \caption{Average five-fold test Dice values of our SE+AW model compared to three other 3D CNNs on KES+MCW and ATLAS datasets. \label{results2}}
\end{figure}

In Figure~\ref{results2} we compare our SE+AW (squeeze-and-excite in fusion and amplified weighting) model to three other 3D CNN models on KES+MCW. On ATLAS we show our ReLUOnly model although the SE+AW variant is only slightly behind in accuracy. On KES+MCW we do not have results for AnatomyNet since that is designed for single modality images. We show the Dice values for four different thresholds of lesion sizes starting from the smallest 25\% to 100\% that includes the entire dataset.

On KES+MCW we see that in the 25\% smallest lesions our model has a 34.4\% accuracy whereas the next best 3D U-Net is 29\%.On ATLAS in the 25\% smallest lesions our model has 41.4\% accuracy while the next best AnatomyNet is 34.1\%. Our weighting techniques (amplified weighting and ReLUOnly) thus give us a 5.4\% and 7\% improvement over the next best on KES+MCW and ATLAS respectively. In the case of ATLAS the improvement is statistically significant with a p-value of 0.004. 

On KES+MCW (100\% threshold) our model has a Dice value of 65.1\% while the next best 3D U-Net reaches 60.8\%. On all of ATLAS our model has Dice value 60.5\% while the next best AnatomyNet reaches 56.1\%. In both cases the differences are strongly statistically significant. On KES+MCW the difference has p-value 0.0016 and on ATLAS in the order of $10^{-6}$. Thus we see that our model not only improves upon previous methods on small hard to detect lesions but overall as well.

In Table~\ref{flops} we see that our model has the fewest FLOPS except for DeepMedic that uses patches of images and thus has a much smaller input size ($19\times19\times19$ and $25\times25\times25$ for DeepMedic vs. $160\times192\times160$ for all other models). Our model total parameters are comparable to 3D-UNet. 
This suggests that our model's performance can be attributed to its architecture and novel design features that we introduce and not just higher capacity.

\begin{table}[h]
\centering
\begin{tabular}{ccccc} \hline
Method & Our 3D CNN model & AnatomyNet & 3D-UNet & DeepMedic \\
Total trainable parameters & 1,739,146 & 1,359,747 & 1,781,304 & 658,295 \\
FLOPS & 1,025,567,394 & 10,765,767,799 & 1,720,811,520 & 91,426,315\\ \hline
\end{tabular}
\caption{Total trainable parameters and FLOPS of our model compared to AnatomyNet, 3D-UNet, and DeepMedic. \label{flops}}
\end{table}

\newpage
\small
\bibliographystyle{unsrt}

\bibliography{my_bib}

\begin{thebibliography}{10}

\bibitem{ronneberger2015u}
Olaf Ronneberger, Philipp Fischer, and Thomas Brox.
\newblock U-net: Convolutional networks for biomedical image segmentation.
\newblock In {\em International Conference on Medical image computing and
  computer-assisted intervention}, pages 234--241. Springer, 2015.

\bibitem{hu2018squeeze}
Jie Hu, Li~Shen, and Gang Sun.
\newblock Squeeze-and-excitation networks.
\newblock In {\em Proceedings of the IEEE conference on computer vision and
  pattern recognition}, pages 7132--7141, 2018.

\bibitem{vaswani2017attention}
Ashish Vaswani, Noam Shazeer, Niki Parmar, Jakob Uszkoreit, Llion Jones,
  Aidan~N Gomez, {\L}ukasz Kaiser, and Illia Polosukhin.
\newblock Attention is all you need.
\newblock In {\em Advances in neural information processing systems}, pages
  5998--6008, 2017.

\bibitem{he2016identity}
Kaiming He, Xiangyu Zhang, Shaoqing Ren, and Jian Sun.
\newblock Identity mappings in deep residual networks.
\newblock In {\em European conference on computer vision}, pages 630--645.
  Springer, 2016.

\bibitem{liew2018large}
Sook-Lei Liew, Julia~M Anglin, Nick~W Banks, Matt Sondag, Kaori~L Ito, Hosung
  Kim, Jennifer Chan, Joyce Ito, Connie Jung, Nima Khoshab, et~al.
\newblock A large, open source dataset of stroke anatomical brain images and
  manual lesion segmentations.
\newblock {\em Scientific data}, 5:180011, 2018.

\bibitem{kamnitsas2017efficient}
Konstantinos Kamnitsas, Christian Ledig, Virginia~FJ Newcombe, Joanna~P
  Simpson, Andrew~D Kane, David~K Menon, Daniel Rueckert, and Ben Glocker.
\newblock Efficient multi-scale 3d cnn with fully connected crf for accurate
  brain lesion segmentation.
\newblock {\em Medical image analysis}, 36:61--78, 2017.

\bibitem{zhu2019anatomynet}
Wentao Zhu, Yufang Huang, Liang Zeng, Xuming Chen, Yong Liu, Zhen Qian, Nan Du,
  Wei Fan, and Xiaohui Xie.
\newblock Anatomynet: Deep learning for fast and fully automated whole-volume
  segmentation of head and neck anatomy.
\newblock {\em Medical physics}, 46(2):576--589, 2019.

\bibitem{he2016deep}
Kaiming He, Xiangyu Zhang, Shaoqing Ren, and Jian Sun.
\newblock Deep residual learning for image recognition.
\newblock In {\em Proceedings of the IEEE conference on computer vision and
  pattern recognition}, pages 770--778, 2016.

\bibitem{zijdenbos1994morphometric}
Alex~P Zijdenbos, Benoit~M Dawant, Richard~A Margolin, and Andrew~C Palmer.
\newblock Morphometric analysis of white matter lesions in mr images: method
  and validation.
\newblock {\em IEEE transactions on medical imaging}, 13(4):716--724, 1994.

\bibitem{wilcoxon1945individual}
Frank Wilcoxon.
\newblock Individual comparisons by ranking methods.
\newblock {\em Biometrics bulletin}, 1(6):80--83, 1945.

\bibitem{dietterich1998approximate}
Thomas~G Dietterich.
\newblock Approximate statistical tests for comparing supervised classification
  learning algorithms.
\newblock {\em Neural computation}, 10(7):1895--1923, 1998.

\bibitem{demvsar2006statistical}
Janez Dem{\v{s}}ar.
\newblock Statistical comparisons of classifiers over multiple data sets.
\newblock {\em Journal of Machine learning research}, 7(Jan):1--30, 2006.

\bibitem{japkowicz2011evaluating}
Nathalie Japkowicz and Mohak Shah.
\newblock {\em Evaluating learning algorithms: a classification perspective}.
\newblock Cambridge University Press, 2011.

\end{thebibliography}

\end{document}